\documentstyle[preprint,aps,epsf]{revtex}
\def\simgt{\,\rlap{\lower 3.5pt\hbox{$\mathchar \sim$}}\raise 1pt\hbox{$>$}\,}
\def\simlt{\,\rlap{\lower 3.5pt\hbox{$\mathchar \sim$}}\raise 1pt\hbox{$<$}\,}
\begin{document}

\draft
\tightenlines

\title{
\vspace*{-35pt}
{\normalsize \hfill {\sf TKYNT-05-16, \ \ 2005/June}} \\
Lee-Yang zero analysis for the study of QCD phase structure
}
\author{Shinji Ejiri}
\address{Department of Physics, The University of Tokyo, 
Tokyo 113-0033, Japan}

\date{\today}
\maketitle

\begin{abstract}
We comment on the Lee-Yang zero analysis for the 
study of the phase structure of QCD at high temperature and baryon 
number density by Monte-Carlo simulations. 
We find that the sign problem for non-zero density QCD 
induces a serious problem in the finite volume scaling analysis 
of the Lee-Yang zeros for the investigation of the order of the 
phase transition. If the sign problem occurs at large volume, 
the Lee-Yang zeros will always approach the real axis of 
the complex parameter plane in the thermodynamic limit. 
This implies that a scaling behavior which would 
suggest a crossover transition will not be obtained. 
To clarify this problem, we discuss the Lee-Yang zero analysis 
for SU(3) pure gauge theory as a simple example without the sign 
problem, and then consider the case of non-zero density QCD. 
It is suggested that the distribution of the Lee-Yang zeros in the complex 
parameter space obtained by each simulation could be more 
important information for the investigation of the critical endpoint 
in the $(T, \mu_q)$ plane than the finite volume scaling behavior. 
\end{abstract}




\section{Introduction}
\label{sec:intro}

In the last few years, remarkable progress in exploring the QCD phase 
structure in the temperature $(T)$ and quark chemical potential 
$(\mu_q)$ plane have been made in numerical studies of lattice QCD. 
The phase transition line, separating hadron phase and quark-gluon 
plasma phase, was investigated from $\mu_q=0$ to finite $\mu_q$ 
\cite{FK1,FK2,FK3,us02,dFP1,dEL1}, 
and the equation of state was also analyzed quantitatively 
at low density \cite{us02,us03,GG1,us05}. 

Among others, study of the endpoint of the first order phase 
transition line in the $(T, \mu_q)$ plane, whose existence is suggested 
by phenomenological studies \cite{AY,SRS}, is particularly important 
both from the experimental and theoretical point of view. 
To locate the critical endpoint, Fodor and Katz\cite{FK2,FK3} investigated 
the positions of the Lee-Yang zeros (to be explained more below) 
in the complex $\beta=6/g^2$ plane using lattices with different 
spatial volumes, and examined the finite-volume scaling behavior 
of a  Lee-Yang zero closest to the real axis.
There are also studies in which the behavior of the critical endpoint 
as a function of the quark masses is examined 
by using the property that a critical endpoint exists at 
$\mu_q=0$ in the very small quark mass region for QCD with three 
flavors having degenerate quark masses. 
For extrapolating the result to the case with physical quark masses, 
an approach on the basis of the Taylor expansion in terms of 
$\mu_q/T$ \cite{crtpt} and that of the imaginary chemical potential 
\cite{dFP2,dEL2} have been developed.
Moreover, a study of phase-quenched finite density QCD, i.e. 
simulations with an isospin chemical potential, 
has been discussed in Ref.~\cite{KS04}.
The radius of convergence in the framework of the Taylor expansion 
of the grand canonical potential can establish a 
lower bound of the location of the critical endpoint \cite{us03,us05,GG2}.
Also, the Glasgow method \cite{Bar} is an interesting approach 
for the study of QCD at non-zero baryon density.

In this paper, we focus our attention on the method of 
the Lee-Yang zero \cite{LeeYang} applied to finite density QCD. 
The Lee-Yang zero analysis is a popular method that is used to investigate the 
order of phase transitions. In order to study the existence of 
singularities in the thermodynamic limit (infinite volume limit), 
Lee and Yang proposed the following approach: 
Because there are no thermodynamic singularities as long as the volume is 
finite, the partition function ${\cal Z}$ is always non-zero; it can develop
zeros only in the 
infinite volume limit. However, if a real parameter of the model is 
extended into the complex parameter plane, a singularity, 
characterized by ${\cal Z}=0$, can appear in the complex parameter plane
even in a finite volume. These zeros are called the Lee-Yang zeros. 
Therefore, one can find the position of a singularity by exploring the 
position of the Lee-Yang zero in the complex parameter plane as a 
function of volume, and extrapolating the position of the Lee-Yang 
zero in the thermodynamic limit. 
For a system with a first order phase transition, the position of 
the nearest Lee-Yang zero approaches to
the real axis in inverse proportion to the volume.
On the other hand, the Lee-Yang zeros do not reach the real axis for crossover 
transition, i.e. rapid change without any thermodynamic singularities. 
For QCD at non-zero baryon density, we expect a rapid crossover 
transition in the low density regime which changes into a first order 
phase transition  beyond a critical value of the density.

As we mentioned, Fodor and Katz have investigated the finite 
volume dependence of the Lee-Yang zeros in the complex $\beta \equiv 6/g^2$ 
plane for various values of the chemical potential. To carry out such analysis, 
the reweighting method \cite{Swen88} is adopted, in which 
one performs simulations at $\mu_q=0$, and then corrects for the modified 
Boltzmann weight in the measurement of observables. 
In this case a famous problem arises for large $\mu_q/T$ and 
large volume, which is called ``sign problem".  
The sign problem is caused by complex phase fluctuations of the 
fermion determinant. In the region of small $\mu_q/T$, the phase 
fluctuations are not large and the sign problem is not serious. However, 
if the sign of the modification factor changes frequently during 
subsequent Monte-Carlo steps for large $\mu_q/T$, the statistical 
error becomes larger than the expectation values in general. 

We find that the sign problem induces a serious problem 
in the finite volume scaling analysis of the Lee-Yang zero, 
which is used by Fodor and Katz. For any non-zero $\mu_q$
the normalized partition function calculated on the real axis 
with necessarily limited statistics will numerically always 
be consistent with zero once the volumes grow large.
This means that the scaling behavior suggesting a crossover transition 
will not be obtained for the case with the sign problem, 
which is in contrast with the usual expectation. 

Before discussing the case of non-zero density QCD, 
we study in Sec.~\ref{sec:pureQCD},
as a simple example, the Lee-Yang zeros in the complex $\beta$ plane for 
SU(3) pure gauge theory  
by analyzing data from Monte-Carlo simulations.  This model has 
a first order phase transition \cite{FOU} and simulations are 
much easier than for QCD at non-zero baryon density. 
Moreover, the pure gauge theory does, of course, not have a sign problem. 
Hence it is a good example to demonstrate how the Lee-Yang analysis works
in the complex $\beta$ plane. In addition, it will become clear
during this exercise that the  
complex phase fluctuations arising from the imaginary part of $\beta$ near 
Lee-Yang zeros are quite similar to those coming from the quark determinant 
where the sign problem exists for non-zero density QCD. 
The problem of the complex measure is reviewed in Sec.~\ref{sec:phase}. 
There we also comment on the reweighting method for the study at 
non-zero baryon density. 
In Sec.~\ref{sec:nonzeromu}, we discuss a problem which arises when 
we apply the Lee-Yang zero analysis for non-zero baryon density QCD by 
using the reweighting technique, and consider possible other approaches 
in the framework of the Lee-Yang zero analysis for the investigation 
of the critical endpoint. Conclusions and discussions are given in 
Sec.~\ref{sec:concl}.

\section{Lee-Yang zero for SU(3) pure gauge theory}
\label{sec:pureQCD}

\subsection{General remarks}
In this section, we apply the method of Lee-Yang zeros to the
SU(3) pure gauge theory (quenched QCD).\footnote{
A pioneering study has been done for lattices with $N_{\tau}=2$ 
in Ref.~\cite{KSC}.}
The phase transition of the SU(3) pure gauge theory is known to be of first 
order \cite{FOU}, which is expected from the corresponding Z(3) spin models. 
The pure gauge theory is controlled by only one parameter $\beta=6/g^2$ 
with the partition function,
\begin{eqnarray}
{\cal Z}= \int {\cal D} U e^{6 \beta N_{\rm site} P}, 
\end{eqnarray}
where $P$ is an averaged plaquette 
$P= (\sum_{x, \mu < \nu} W_{\mu \nu}^{1 \times 1}(x))/(6N_{\rm site})$, 
and $W_{\mu \nu}^{1 \times 1}$ is $1 \times 1$ Wilson loop operator for 
the lattice size $N_{\rm site}=N_{\sigma}^3 \times N_{\tau}$. 
Here, $N_{\sigma}$ and $N_{\tau}$ are spatial and temporal extension of 
the lattice. We extend the real parameter $\beta$ into the complex plane
$(\beta_{\rm Re}, \beta_{\rm Im})$, and determine 
the position of Lee-Yang zeros, at which 
${\cal Z}(\beta_{\rm Re}, \beta_{\rm Im})=0$ is satisfied, 
by numerical simulations. 
We use standard Monte-Carlo techniques; 
configurations $\{ U_{\mu} \}$ are generated 
with the probability of the Boltzmann weight. The expectation 
value of an operator ${\cal O}[U_{\mu}]$, 
$\left\langle {\cal O} \right\rangle$, is then calculated by taking an 
average over the configurations. 
We expect that the position of the Lee-Yang zero 
$(\beta_{\rm Re}^0, \beta_{\rm Im}^0)$ approaches the real $\beta$ 
axis in the infinite volume limit; 
with $\beta_{\rm Im}^0 \sim 1/V \equiv N_{\sigma}^{-3}$ 
for a first order phase transition.

In carrying out the above calculation two problems arise:
One is that the Monte Carlo method is applicable only to the expectation 
values of physical quantities but not to the partition function itself. 
Another problem is that the measure is complex for a complex coupling
$\beta$, and hence we cannot apply the Monte Carlo method directly, since 
the probabilities (Boltzmann weights) must be real and positive.
To avoid these problems, we introduce the normalized partition function 
${\cal Z}_{\rm norm}$ together with the reweighting technique, 
\begin{eqnarray}
{\cal Z}_{\rm norm}(\beta_{\rm Re}, \beta_{\rm Im})
& \equiv & \left| \frac{{\cal Z}(\beta_{\rm Re}, \beta_{\rm Im})}
{{\cal Z}(\beta_{\rm Re}, 0)} \right|
= \left| 
\frac{\int {\cal D} U e^{6(\beta_{\rm Re} +i\beta_{\rm Im}) N_{\rm site} P}} 
{\int {\cal D} U e^{6\beta_{\rm Re} N_{\rm site} P}} \right|
\nonumber \\
&=& \left| \left\langle e^{6i\beta_{\rm Im} N_{\rm site} P}
 \right\rangle_{(\beta_{\rm Re}, 0)} \right|
= \left| \left\langle e^{6i\beta_{\rm Im} N_{\rm site} \Delta P}
 \right\rangle_{(\beta_{\rm Re}, 0)} \right| . 
\label{eq:znorm} 
\end{eqnarray}
Here $\Delta P = P - \langle P \rangle$ and 
$ \left| \exp(6i\beta_{\rm Im} N_{\rm site} \left\langle P 
\right\rangle ) \right| =1$. 
Because the denominator ${\cal Z}(\beta_{\rm Re}, 0)$ is always finite 
for any finite volume, the position of 
${\cal Z}(\beta_{\rm Re}, \beta_{\rm Im})=0$ can be identified by 
analyzing ${\cal Z}_{\rm norm}$. 
Although the partition function is not zero for $\beta_{\rm Im}=0$, 
it can be zero at some points in the $(\beta_{\rm Re}, \beta_{\rm Im})$ 
plane, when the complex phase factor in Eq.(\ref{eq:znorm}) changes 
sign frequently on the generated configurations.
For the determination of the critical point in the original theory, 
i.e. on the real $\beta$ axis, the position of the nearest Lee-Yang 
zero should be investigated as a function of the volume $V=N_{\sigma}^3$.

The mechanism that leads the occurrence of a Lee-Yang zero in 
Eq.(\ref{eq:znorm}) is quite similar to that which limits the 
applicability of the reweighting method for QCD with finite 
chemical potential \cite{us02,eji04}.
We will discuss this in more detail in Sec.~\ref{sec:phase}. 
At a point for which the width of the probability distribution 
of $6\beta_{\rm Im} N_{\rm site} P$ is smaller than $O(\pi/2)$, 
the sign of the complex phase does not change. 
Therefore, the standard deviation of the plaquette distribution is 
required to be larger than $\pi/(12 \beta_{\rm Im} N_{\rm site})$ 
in the region where Lee-Yang zeros exist.
Moreover, because the square of standard deviation is in proportion to 
the value of the plaquette susceptibility, the position of its maximum 
must agree with the position where $Z_{\rm norm}$ becomes minimal as
function of $\beta_{\rm Re}$ for fixed $\beta_{\rm Im}$. 
Hence, the real part, $\beta_{\rm Re}$, of the position of the nearest 
Lee-Yang zero must be consistent with the peak 
position of the plaquette susceptibility. The method to find a 
critical point from the location of Lee-Yang zeros
thus is essentially the same as the method which determines a 
critical point through the location of the peak position of the susceptibility
and its finite volume scaling. 

Here, it is instructive to introduce a probability distribution function 
for the plaquette, $w(P)$, which is defined by 
\begin{eqnarray}
w(P') =\frac{1}{\cal Z} \int {\cal D} U \delta(P'-P)
e^{6\beta_{\rm Re} N_{\rm site} P}, 
\label{eq:pdist}
\end{eqnarray}
where $\delta(x)$ is the delta function. 
Then, Eq.(\ref{eq:znorm}) can be rewritten as 
\begin{eqnarray}
{\cal Z}_{\rm norm}(\beta) = \left| \int {\rm d} P
e^{6i \beta_{\rm Im} N_{\rm site} \Delta P} w(P) \right| .
\label{eq:zdist}
\end{eqnarray}
This means that the partition function ${\cal Z}_{\rm norm}$ as a 
function of $6 \beta_{\rm Im} N_{\rm site}$ is obtained through a Fourier 
transformation of $w(P)$. 

Using this equation, the relation between the scaling behavior of 
the Lee-Yang zeros in the infinite volume limit and the 
distribution function of the plaquette becomes clearer. 
In Monte-Carlo simulations, configurations with probabilities proportional
to their Boltzmann weight are generated by a computer, and we 
obtain a distribution function of the plaquette from 
the histogram of the plaquette. 
The histogram has usually a Gaussian shape at a normal, non-critical point, 
but it deviates from the Gaussian form near a critical point, 
and attains a double peak shape at a first order transition point, 
corresponding to the coexistence of two phases. 

For the case of a non-singular point of $\beta_{\rm Re}$ or 
a crossover pseudo-critical point, where
the distribution is expected to be a Gaussian function, 
the point of $Z_{\rm norm}=0$ does not exist except in the limit of
$\beta_{\rm Im} N_{\rm site} \to \infty$ or $-\infty$, because the 
function which is obtained through a Fourier transformation of a Gaussian 
function again is a Gaussian function. Of course, 
results of numerical simulations have statistical errors, 
hence ${\cal Z}_{\rm norm}$ can become zero ``within errors'', 
if the expectation value 
and the error become of the same order. However, in this case, the point  
at which $Z_{\rm norm}=0$ appears at random in terms of 
$\beta_{\rm Im} \times N_{\rm site}$. Therefore, the volume dependence 
of the position of the Lee-Yang zero $(\beta_{\rm Re}^0, \beta_{\rm Im}^0)$ 
does not necessary to be 
$\beta_{\rm Im}^0 \sim 1/V (\equiv N_{\sigma}^{-3})$ for fixed $N_{\tau}$.

On the other hand, in the case of a first order phase transition, 
we expect that the plaquette histogram has two peaks having 
the same peak height at the transition point. 
Performing the Fourier transformation of such a double peaked function 
leads to a function which has zeros periodically. 
For example, a distribution function $w(P)$ having two Gaussian 
peaks at $\Delta P = \pm A$ leads to a normalized partition function 
${\cal Z}_{\rm norm}$ which has zeros at 
\begin{eqnarray}
\beta_{\rm Im}^0 = \frac{\pi (2n+1)}{12N_{\rm site} A},
\hspace{5mm} (n=0,1,2,3,\cdots).
\label{eq:zerowp}
\end{eqnarray}
This is mathematically the same as that for the interference experiment  
using a laser and a double-slit. The Lee-Yang zeros correspond to 
dark lines (destructive interference) and they appear periodically as 
given in Eq.(\ref{eq:zerowp}).
Moreover, for a first order phase transition the difference of plaquette 
values in cold and hot phases, 
$2A$, is related to the latent heat $\Delta \varepsilon$, 
i.e. the energy difference between the hot and cold phases, 
\begin{eqnarray}
\frac{\Delta \varepsilon}{T^4} \approx -12A N_{\tau}^4 a\frac{d\beta}{da}, 
\label{eq:de}
\end{eqnarray}
where $a$ is the lattice spacing.
Since $\Delta \varepsilon$ is non-zero, $A$ does not vanish in 
the infinite volume limit $(N_{\sigma}^3 \to \infty)$. 
Therefore, we find that in the infinite volume limit the nearest Lee-Yang zero 
approaches the real $\beta$ axis like $\beta_{\rm Im}^0 \sim 1/V$,
which is consistent with the general 
argument on the Lee-Yang zero for a first order phase transition. 
We also emphasize that the isolated Lee-Yang zeros appear periodically.
The distances to these points from the real axis are $1, 3, 5, \cdots$ 
in units of the distance to the first Lee-Yang zero. This is also 
an important property, which is not observed for a crossover transition.

In addition, the discussion given for the plaquette distribution function 
can be extended to the analysis of fourth order Binder cumulants, 
\begin{eqnarray}
B_4 \equiv \frac{\left\langle \Delta P^4 \right\rangle}
{\left\langle \Delta P^2 \right\rangle^2}, 
\label{eq:binder}
\end{eqnarray}
which is an alternative to the method of Lee-Yang zeros
often used to identify the order of a phase transition.
The value of the Binder cumulant at the critical point depends on 
the universality class. 
In the case of a first order phase transition, 
assuming the plaquette distribution is a double peaked function, 
the Binder cumulants are estimated as 
\begin{eqnarray}
B_4 = \frac{\int {\rm d} P \Delta P^4 w(P)}
{\left( \int {\rm d} P \Delta P^2 w(P) \right)^2}
\approx \frac{A^4}{(A^2)^2} \approx 1, 
\label{eq:b4dp}
\end{eqnarray}
where the distance between two peaks is $2A$ and is wider 
than the width of each peak. 
On the other hand, when the distribution function can be modeled 
by a Gaussian function for a crossover transition, 
the Binder cumulants are given by 
\begin{eqnarray}
B_4 \approx \frac{\sqrt{x/\pi} \int dP \Delta P^4 e^{-x \Delta P^2}}
{\left(\sqrt{x/\pi}\int dP \Delta P^2 e^{-x \Delta P^2} \right)^2}
= \left(\sqrt{\frac{x}{\pi}} \frac{d^2 \sqrt{\pi/x}}{d x^2}
\right) \left/ \left(- \sqrt{\frac{x}{\pi}} 
\frac{d \sqrt{\pi/x}}{d x} \right)^2 \right.
= 3.
\label{eq:b4dp2}
\end{eqnarray}
In a region where a first order phase transition changes to 
a crossover, the Binder cumulant changes rapidly from one to three. 
We expect to find such a region for full QCD at high temperature and density. 
The value of the Binder cumulant at the endpoint of the first 
order transition line, which is of second order, is determined 
by the universality class. Hence, the plaquette distribution 
function plays an important role for both methods to 
identify the order of a phase transition.

\subsection{Numerical results}

We calculate the normalized partition function for SU(3) 
pure gauge theory to find Lee-Yang zeros in the complex 
$\beta$ plane, using data for plaquettes obtained by QCDPAX 
in Ref.~\cite{QCDPAX}. 
There are five data sets measured at the transition point for 
$N_{\tau}=4$ and $6$. The spatial lattice sizes are
$24^2 \times 36 \times 4, 12^2 \times 24 \times 4, 
36^2 \times 48 \times 6, 24^3 \times 6,$ and $20^3 \times 6$.
$O(10^6)$ configurations are available for the analysis of each 
data set. 
The reweighting technique is also used for the real $\beta$ direction  
to analyze the Lee-Yang zeros in the complex $\beta$ 
plane for a data set obtained at only one $\beta$ point 
$(\beta_{\rm Re}, \beta_{\rm Im})=(\beta_0, 0).$ 
The normalized partition function is given by 
\begin{eqnarray}
{\cal Z}_{\rm norm}(\beta_{\rm Re}, \beta_{\rm Im}) 
= \left| \frac{\left\langle e^{6i\beta_{\rm Im} N_{\rm site} \Delta P} 
e^{6\beta_{\rm Re} N_{\rm site} \Delta P} 
\right\rangle_{(\beta_0, 0)}}
{\left\langle e^{6\beta_{\rm Re} N_{\rm site} \Delta P}
 \right\rangle_{(\beta_0, 0)}}
\right| . 
\label{eq:nzrew}
\end{eqnarray}

Figure \ref{fig:lyznt4} shows the contour plot of ${\cal Z}_{\rm norm}$ 
for the $24^2 \times 36 \times 4$ lattice. 
The simulation point is $\beta_0=3.6492$. 
In this definition, ${\cal Z}_{\rm norm}$ is normalized to be one on 
the real $\beta$ axis. Circles at 
$(\beta_{\rm Re}^0, \beta_{\rm Im}^0) \approx (5.6925, 0.0021)$ and 
$(5.6931, 0.0056)$ are Lee-Yang zeros.
Since the SU(3) pure gauge theory has a first order phase transition, 
Lee-Yang zeros appear periodically.
For this data, two clear peaks are visible in the plaquette histogram 
\cite{QCDPAX}. The distance between these two peaks is $2A \approx 0.003$. 
The positions of the Lee-Yang zeros are consistent with 
$\beta_{\rm Im}^0 \sim \pi / (12N_{\rm site}A) \approx 0.002$ and 
$3 \pi / (12N_{\rm site}A) \approx 0.006$, as given in Eq.(\ref{eq:zerowp}). 

\begin{figure}[t]
\centerline{
\epsfxsize=10.0cm\epsfbox{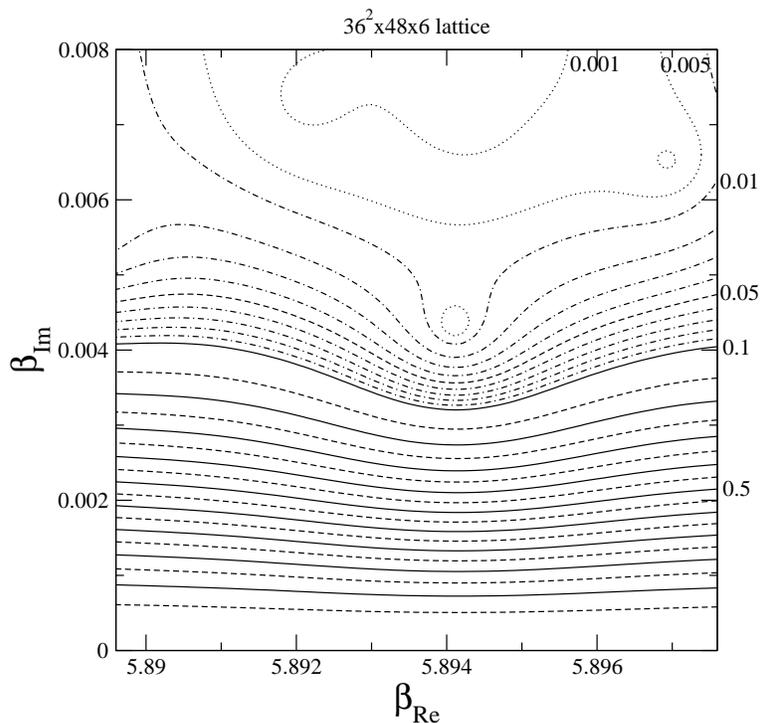}
}
\caption{
Contour plot of the normalized partition function 
${\cal Z}_{\rm norm}$ for the $36^2 \times 48 \times 6$ lattice. 
The simulation point is $\beta_0=5.8936$.
}
\vspace*{-4mm}
\label{fig:lyznt6}
\end{figure}

Above property is not seen so clearly for lattices having small 
$N_{\sigma}$ and large $N_{\tau}$. The position of
the next-to-leading zero points of 
${\cal Z}_{\rm norm}$ appear at random for the other data sets 
relative to the nearest zero point. 
The positions of Lee-Yang zeros are shown in Table~\ref{tab:flyz}. 
We could not obtain clearly isolated Lee-Yang zeros for the lattices 
$24^3 \times 6$ and $20^3 \times 6$. The second nearest Lee-Yang 
zero to the real axis could be measured only for the 
$24^2 \times 36 \times 4$ lattice. 
The result on the $36^2 \times 48 \times 6$ lattice $(\beta_0=5.8936)$ 
is also shown in Fig.~\ref{fig:lyznt6}. Only the nearest Lee-Yang zero 
is obtained clearly. The Lee-Yang zero becomes less clear as 
$N_{\tau}$ increases and $N_{\sigma}$ decreases, hence 
simulations on lattices having large $N_{\sigma}/N_{\tau}$ seem 
to be necessary for the study of Lee-Yang zeros.

The values for $\beta_{\rm Im}^0 V$ on 
$24^2 \times 36 \times 4$ and $12^2 \times 24 \times 4$ lattices 
are $43.9(5)$ and $42.0(6)$, respectively. These are roughly constant 
and suggest the scaling behavior of $\beta_{\rm Im}^0 \sim 1/V$ for a 
first order phase transition. 
Also, in the previous study for lattices with $N_{\tau}=2$ \cite{KSC}, 
the $1/V$ scaling behavior has been confirmed for 
$N_{\sigma}=6, 8, 10$ and $12$.
However, for a more precise quantitative investigation that takes into 
account the errors, the spatial lattice size $12^2 \times 24$ may not be 
large enough to check the $1/V$ scaling for $N_{\tau}=4$, since the 
difference of $\beta_{\rm Im}^0 V$ is larger than the statistical error. 
We should fit the data obtained on more than two lattices 
by a curved function of $1/V$, to confirm through a $1/V$ scaling analysis
that the phase transition of the SU(3) pure gauge theory is first order. 
E.g. for the study of the SU(2) gauge-Higgs 
model \cite{yaoki}, the following fitting functions have been used, 
${\rm Im} \kappa_0(V)=\kappa_0^c+CV^{-\nu}$ and
${\rm Im} \kappa_0(V)=\kappa_0^c+CV^{-1}+DV^{-2}$ for a complex 
parameter $\kappa$ in the model with fitting parameters 
$\kappa_0^c, C, D,$ and $\nu$.

\section{Complex phase fluctuation and sign problem}
\label{sec:phase}

As seen in the previous section, the investigation of Lee-Yang 
zeros in the complex $\beta$ plane seems to be useful to identify 
the order of phase transition. 
However, if we try to extend this analysis to full QCD 
at non-zero baryon density, a serious problem arises. 
This problem is closely related to the sign problem for 
finite density QCD, since the normalized partition function 
can be zero in the complex $\beta$ plane due to fluctuations of 
the complex phase related to $\beta_{\rm Im}$
and also due to the complex phase from the quark determinant that
causes the sign problem.
Before discussing the Lee-Yang zero analysis for finite density 
QCD, we would like to review the sign problem briefly. 

The main difficulty for studies at finite baryon density is that 
the Boltzmann weight is complex if the chemical potential is non-zero. 
In this case the Monte-Carlo method is not applicable directly, since 
configurations cannot be generated with a complex probability.
One approach to avoid this problem is the reweighting method. 
We perform simulations at $\mu=0$, and incorporate the remaining
part of the correct Boltzmann weight for finite $\mu$ in the calculation
of expectation values. 
Expectation values $\langle {\cal O} \rangle$ at $(\beta, \mu)$ are thus
computed by a simulation at $(\beta, 0)$ using the following identity, 
\begin{eqnarray}
\langle {\cal O} \rangle_{(\beta, \mu)} 
= \frac{\left\langle {\cal O} 
e^{N_{\rm f} (\ln \det M(\mu) - \ln \det M(0))}
\right\rangle_{(\beta,0)}}{ \left\langle
e^{N_{\rm f} (\ln \det M(\mu) - \ln \det M(0))}
\right\rangle_{(\beta,0)}}, 
\label{eq:murew}
\end{eqnarray} 
where $M$ is the quark matrix and $N_{\rm f}$ is the number of flavors 
($N_{\rm f}/4$ for staggered type quarks instead of $N_{\rm f}$); 
$\mu$ is a quark chemical potential in lattice units, i.e. 
$\mu \equiv \mu_q a=\mu_q/(N_{\tau} T)$, and $\mu_q$ is the quark 
chemical potential in  physical units.
This is the basic formula of the reweighting method. 
However, because $\ln \det M(\mu)$ is complex, 
the calculations of the numerator and denominator in Eq.(\ref{eq:murew}) 
becomes in practice increasingly more difficult for larger $\mu$. 
We define the phase of the quark determinant $\theta$ by 
$(\det M(\mu))^{N_{\rm f}/4} \equiv |\det M(\mu)|^{N_{\rm f}/4} 
e^{i \theta}$ for staggered type quarks. 
If the typical value of $\theta$ becomes larger than $\pi/2$, the real 
part of $e^{i \theta}$ $(=\cos \theta)$ changes its sign frequently.
Eventually both the numerator and denominator of Eq.(\ref{eq:murew}) become 
smaller than their statistical errors and Eq.(\ref{eq:murew}) 
can no longer be evaluated. We call it the ``sign problem''.

Here, the denominator of Eq.(\ref{eq:murew}), or simply 
$\langle \cos \theta \rangle$, is a good indicator for the occurrence of
the sign problem. If this indicator is zero within statistical errors, 
Eq.(\ref{eq:murew}) cannot be computed. In the following we give an 
estimate for the value of the complex phase.
Since the direct calculation of the quark determinant is difficult 
except for calculations on small lattices, we expand 
$\ln \det M(\mu)$ in a Taylor series, 
\begin{eqnarray}
\ln \det M(\mu) - \ln \det M(0) = \sum_{n=1}^{\infty} 
\left[ \frac{\partial^n (\ln \det M)}{\partial \mu^n} \right] 
\frac{\mu^n}{n!} . 
\end{eqnarray}
Then, we can easily separate it into real and imaginary parts 
because the even derivatives of $\ln \det M(\mu)$ are real and 
the odd derivatives are purely imaginary \cite{us02}. 
The complex phase $\theta$ is given by 
\begin{eqnarray}
\theta = \frac{N_{\rm f}}{4} \sum_{n=1}^{\infty} 
{\rm Im} \frac{\partial^{2n-1} (\ln \det M)}{\partial \mu^{2n-1}} 
\frac{\mu^{2n-1}}{(2n-1)!} , 
\label{eq:theta}
\end{eqnarray} 
for staggered type quarks at small $\mu$.  
The Taylor expansion coefficients are rather easy to calculate 
by using the stochastic noise method. 
The comparison between the value of $\theta$ with this approximation and 
the exact value has been done in Ref.~\cite{dFKT}, and the reliability 
and the application range have been discussed.

\begin{figure}[t]
\centerline{
\epsfxsize=10.0cm\epsfbox{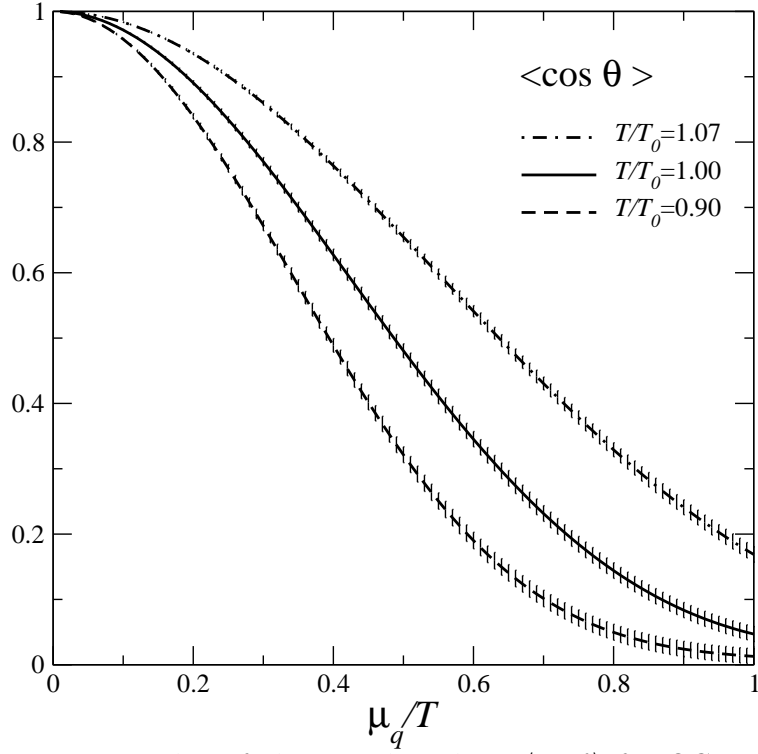}
}
\caption{
The expectation value of the complex phase $\langle \cos \theta \rangle$ 
for QCD with two flavors of p4-improved staggered quarks at $ma=0.1$.
}
\vspace*{-4mm}
\label{fig:phind}
\end{figure}

We use data for the Taylor expansion coefficients obtained in 
Ref.~\cite{us05}. The data were generated by using Symanzik-improved 
gauge and p4-improved staggered fermion actions. 
Coefficients up to $O(\mu^5)$ have been calculated. 
Figure \ref{fig:phind} shows the indicator $\langle \cos \theta \rangle$ 
measured at $\beta=\{3,60, 3.65,$ and $3.68\}$, for $ma=0.1$, 
corresponding to $T/T_c= 0.90, 1.00,$ and $1.07$, respectively, 
on a $16^3 \times 4$ lattice\footnote{
As mentioned in Ref.~\cite{us02}, in the calculation using 
the stochastic noise method, the error due to the finite number 
of noise vector $(N_{\rm noise})$ is large for the calculation 
of $\langle \cos \theta \rangle$ with $N_{\rm noise}=10$. 
For the purpose of this study we increased the number of noise vector 
to $N_{\rm noise}=100$. We checked that the difference between 
the results with $N_{\rm noise}=50$ and $N_{\rm noise}=100$ is 
about $10\%$ for the calculation of the position at which 
$\langle \cos \theta \rangle =0.1$.
}. 
We also estimate the values of $\mu_q/T \equiv N_{\tau} \mu$ at which 
$\langle \cos \theta \rangle=0.1, 0.01$ and $0.0$. Results are given 
in Table \ref{tab:mu}.
The situation now is quite similar to the non-singular case of the normalized 
partition function in the complex $\beta$ plane 
discussed in the previous section. This becomes even more apparent
if we consider for simplicity only the first term in the expansion 
of $\theta$ which is proportional to $\mu$. Then
$\langle \cos \theta \rangle$, 
$\mu$ and ${\rm Im} [\partial (\ln \det M)/\partial \mu]$  
correspond to the normalized partition function, $\beta_{\rm Im}$ 
and plaquette, respectively. 

Because the distribution of the complex phase $\theta$ is almost of
Gaussian shape, the indicator, $\langle \cos \theta \rangle$, 
decreases exponentially as $\mu$ increases, 
and it may cross zero at a point where the expectation value 
becomes smaller than the statistical error. 
Therefore, the points of $\langle \cos \theta \rangle =0$ appear 
accidentally and the results given in Table \ref{tab:mu} are unstable.
Moreover, $\partial (\ln \det M)/\partial \mu$ becomes larger as the 
volume increases, hence the indicator for the sign problem vanishes 
in the infinite volume limit for any non-zero $\mu$, which means that 
the range of applicability for the reweighting method approaches $\mu=0$ 
in the infinite volume limit.

\section{Lee-Yang zero analysis for finite density QCD}
\label{sec:nonzeromu}

The Lee-Yang zero analysis for finite density QCD has been 
performed by Fodor and Katz \cite{FK2,FK3}.
They measured the normalized partition function ${\cal Z}_{\rm norm}$ 
using the reweighting method, and determined the points where  
${\cal Z}_{\rm norm}=0$ as a function of spatial volume. 
The normalized partition function is defined by 
\begin{eqnarray}
{\cal Z}_{\rm norm}(\beta_{\rm Re}, \beta_{\rm Im}, \mu)
& \equiv & \left| \frac{{\cal Z}(\beta_{\rm Re}, \beta_{\rm Im}, \mu)}
{{\cal Z}(\beta_{\rm Re}, 0, 0)} \right|
\nonumber \\
&=& \left| 
\left\langle e^{6i\beta_{\rm Im} N_{\rm site} \Delta P} 
e^{i \theta}  
\left| e^{(N_{\rm f}/4) (\ln \det M(\mu) - \ln \det M(0))} \right| 
\right\rangle_{(\beta_{\rm Re}, 0, 0)} \right|
\label{eq:znorm1}
\end{eqnarray} 
or
\begin{eqnarray}
{\cal Z}_{\rm norm}(\beta_{\rm Re}, \beta_{\rm Im}, \mu)
& \equiv & \left| \frac{{\cal Z}(\beta_{\rm Re}, \beta_{\rm Im}, \mu)}
{{\cal Z}(\beta_{\rm Re}, 0, \mu)} \right|
\nonumber \\
&=& \left| \frac{ \left\langle 
e^{6i\beta_{\rm Im} N_{\rm site} \Delta P} e^{i \theta}  
\left| e^{(N_{\rm f}/4) (\ln \det M(\mu) - \ln \det M(0))} \right| 
\right\rangle_{(\beta_{\rm Re}, 0, 0)} }{
\left\langle e^{i \theta}  
\left| e^{(N_{\rm f}/4) (\ln \det M(\mu) - \ln \det M(0))} \right| 
\right\rangle_{(\beta_{\rm Re}, 0, 0)}} \right|
\label{eq:znorm2}
\end{eqnarray} 
for staggered type quarks. $\theta$ is the complex phase of 
$\exp[(N_{\rm f}/4) (\ln \det M(\mu) - \ln \det M(0))]$.
Since the numerator of Eq.(\ref{eq:znorm2}), which is 
the same as ${\cal Z}_{\rm norm}$ in Eq.(\ref{eq:znorm1}), 
is required to be zero at a zero point of 
${\cal Z}_{\rm norm}$ in Eq.(\ref{eq:znorm2}), 
we consider Eq.(\ref{eq:znorm1}) as an indicator for the Lee-Yang zero. 

Here, we notice that for $\beta_{\rm Im}=0$ 
this normalized partition function  
is exactly the same as the indicator for the 
sign problem, i.e. the denominator of Eq.(\ref{eq:murew}). 
As discussed in the previous section, in any practical simulation
this indicator will be consistent with zero within errors for large 
values of $\mu$.  
Moreover, the region where the indicator is non-zero becomes 
narrower as the volume increases, and this region vanishes in the 
infinite volume limit. Hence, the Lee-Yang zeros always approach 
the $\beta_{\rm Im}=0$ axis in the infinite volume limit for any 
finite $\mu$. This means that the scaling behavior for crossover 
will not be obtained for the case with the sign problem. 
This is clearly different from the usual expectation for the QCD 
phase diagram in the $(T, \mu)$ plane. Most model calculations 
suggest that the transition is crossover in the low density region.
This might be a problem of the definition of the normalized 
partition function, Eq.(\ref{eq:znorm1}). 
The normalized partition function on the real $\beta$ axis is 
exactly one for $\mu=0$, but it vanishes for finite $\mu$ in 
the infinite volume limit. 
Therefore, it is very difficult to distinguish the first order 
transition and crossover by investigating the position of 
Lee-Yang zeros as a function of spatial volume. 
This is the most important difference between the definitions 
for the pure gauge theory in Sec.~\ref{sec:pureQCD} and QCD at 
non-zero density. 

The critical endpoint is shown to be located at 
$\mu_B=3\mu_q=725(35) {\rm MeV}$ in Ref.~\cite{FK2} and 
$360(40) {\rm MeV}$ in Ref.~\cite{FK3}, which is inconsistent with 
the above argument. 
This may be a problem of the fitting function. In
Refs.~\cite{FK2,FK3}  the position of Lee-Yang zeros has been fitted by 
$\beta_{\rm Im}^0=A(1/V)+\beta_{\rm Im}^{\infty}$, where 
$A$ and $\beta_{\rm Im}^{\infty}$ are fitting parameters. 
The first order transition and crossover have then been
distinguished by the value of $\beta_{\rm Im}^{\infty}$.
As we discussed in Sec.~\ref{sec:pureQCD}, this fitting function 
is too simple to fit the data of Lee-Yang zeros obtained on 
lattices as small as those used in Ref.~\cite{FK2,FK3}, i.e. $V \le 12^3$.
In fact, if one assumes a curved extrapolation function, all data 
in Table 1 of Ref.~\cite{FK3} seems to approach $\beta_{\rm Im}=0$ 
in the $1/V\rightarrow 0$ limit.

In our argument, the statistical error of $Z_{\rm norm}$, which is
controlled by the number of configurations in the Monte-Carlo simulation,
plays an important role. 
If the statistical error of $Z_{\rm norm}$ becomes much smaller than 
the mean value of $Z_{\rm norm}$ by increasing the statistics 
for each simulation, the method in Ref.~\cite{FK2,FK3} would be applicable.
However, one cannot satisfy this condition in general 
simply because we are looking for the Lee-Yang zero which gives 
$Z_{\rm norm}=0$. Namely, statistical error cannot be smaller than 0. 
Moreover, if the error of $Z_{\rm norm}$ is sizeable, 
there appear fake Lee-Yang zeros which are located 
even closer to the real $\beta$ axis than the true zero 
in the region where the mean value of $Z_{\rm norm}$ is smaller than 
the error. Since we adopt the closest zero as the Lee-Yang zero in 
the actual scaling analysis, we may thus misidentify the true zero 
by the fake one.

\begin{figure}[t]
\centerline{
\epsfxsize=10.0cm\epsfbox{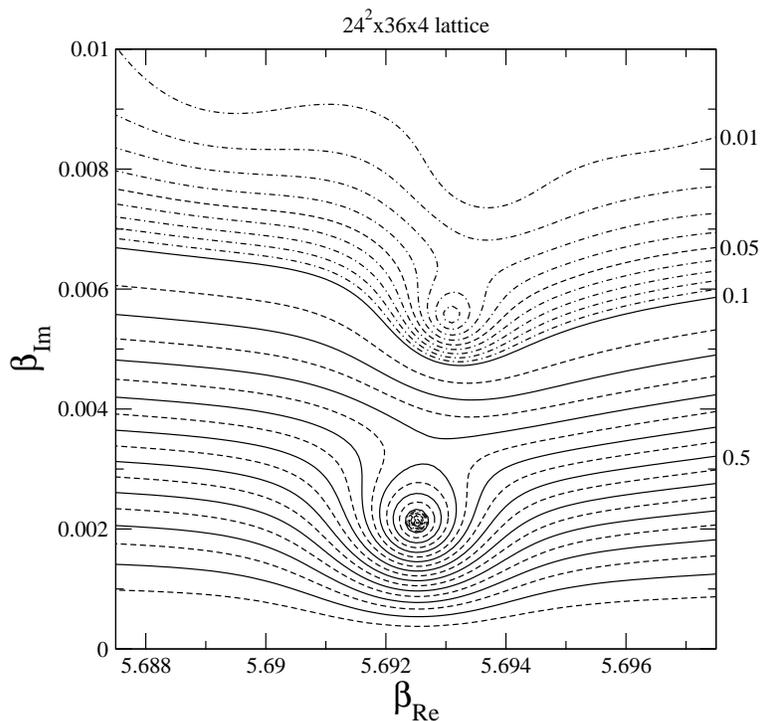}
}
\caption{
Contour plot of the normalized partition function 
${\cal Z}_{\rm norm}$ in the $(\beta_{\rm Re}, \beta_{\rm Im})$ 
plane measured on the $24^2 \times 36 \times 4$ lattice. 
Values in the right edge are ${\cal Z}_{\rm norm}$ . 
The simulation point is $\beta_0=5.6925$.
}
\vspace*{-4mm}
\label{fig:lyznt4}
\end{figure}

The above point can be seen explicitly for the second Lee-Yang zero of 
the SU(3) gauge theory with $N_{\tau}=6$ shown in Fig.~\ref{fig:lyznt6}. 
Theoretically, we expect the second Lee-Yang zero exists around 
$\beta_{\rm Im} \approx 0.013$, i.e. three times larger than that 
of the first Lee-Yang zero as shown in Fig.~\ref{fig:lyznt4}. 
However we find several $\beta$ which give $Z_{\rm norm}=0$ in 
the region of $\beta_{\rm Im} > 0.006$, and they distribute randomly. 
If we identify the second nearest point as the second Lee-Yang zero, 
the resulting $\beta_{\rm Im}$ is much smaller than the theoretical 
expectation. 
This problem is caused by the existence of the region in the complex 
$\beta$-plane where the statistical error of $Z_{\rm norm}$ is larger 
than the mean value as discussed above.
As the statistics is increased for fixed $V$, such a region should 
become smaller and fake Lee-Yang zeros should disappear. 

Now we discuss how the above situation changes by increasing 
the volume $V$. Fortunately, in the SU(3) gauge theory, the location 
of Lee-Yang zero can be determined better also as the volume increases 
as shown in Sec.~\ref{sec:pureQCD}. 
In this case, the validity of the scaling assumption becomes better 
as the volume increases. 
On the other hand, for the case with the sign problem, the error 
normalized by the mean value grows exponentially as a function of 
volume. Then the size of the region having fake Lee-Yang zeros 
cannot be made smaller unless one has exponentially large statistics. 
This leads to the conclusion that the quality of the scaling analysis 
is not improved by increasing the volume, and any reliable information 
about Lee-Yang zeros in the infinite volume limit cannot be obtained. 
This is the reason why the serious problem of identifying the
critical endpoint is intimately related to the sign problem.

This discussion suggests that the finite volume scaling analysis 
suffers serious damage through the unsolved sign problem, and it is very 
difficult to apply the criterion used by Fodor and Katz for the 
investigation of the critical endpoint in practice. 
However, the property of the second nearest Lee-Yang 
zero characteristic for a first order transition in Fig.~\ref{fig:lyznt4}, 
i.e. the fact that the distance to the second Lee-Yang zero from the real 
axis is 
three times larger than that of the first Lee-Yang zero, could be 
investigated on a finite lattice. This study is possible for small 
$\mu$ without taking the infinite volume limit. 
On the other hand, we do not expect any isolated Lee-Yang zero 
for a crossover transition, hence we may be able to determine 
the order of phase transition by investigating the distribution 
of Lee-Yang zeros in the complex $\beta$ plane. 
Although the measurements of the second Lee-Yang zero may require 
large lattice sizes and high statistics, as seen for the case of 
pure SU(3) gauge theory, it may be possible to find the region of 
the first order phase transition, if the critical endpoint exists 
in the low density region.

\section{Conclusions}
\label{sec:concl}

We commented on the Lee-Yang zero analysis for the study of the 
critical endpoint in the $(T, \mu_q)$ phase diagram. 
It is found that the Lee-Yang zero analysis at non-zero baryon density 
encounters a serious problem. 
The complex phases of the quark determinant and the complex $\beta$ 
are mixed at non-zero chemical potential. In this case, in practical 
simulations with limited statistics 
the normalized partition function can develop zeros even on the real 
$\beta$ axis for large $\mu_q$ in finite volumes.
Moreover, in the infinite volume the normalized partition 
function is always zero except for $\mu_q=0$. 
This means that the nearest Lee-Yang zero always approaches 
the real $\beta$ axis in the infinite volume limit. 
The scaling behavior suggesting a crossover transition thus 
will not be obtained. This is clearly different 
from usual expectations for the QCD phase diagram. 
To avoid this problem, the sign problem must be removed by careful 
treatments increasing the number of configurations exponentially as 
the volume or $\mu_q/T$ increases, otherwise the finite volume 
scaling behavior for the position of Lee-Yang zeros, 
which has been analyzed by Fodor and Katz \cite{FK2,FK3}, 
does not provide an appropriate criterion for the investigation 
of the order of the phase transition.

To make the underlying problem more transparent, 
we applied the Lee-Yang zero analysis to the SU(3) pure gauge theory, 
which does not have a sign problem and for which the simulations are 
much easier.  Lee-Yang zeros are found in the complex $\beta$ plane. 
They appear periodically as expected by the discussion using a 
plaquette distribution function for a first order phase transition. 
The positions of the first Lee-Yang zero on two lattices having different 
volume sizes are roughly consistent with the finite size scaling behavior 
for a first order phase transition, i.e. $\beta_{\rm Im}^0 \sim 1/V$. 
However, for quantitative analysis it is necessary to fit data from more than 
two different lattice sizes by a curved function 
to study the order of the phase transition.
It is found, in this analysis, that complex phase fluctuations arising from 
the imaginary part of $\beta$ play an important role, and the mechanism 
that leads to the appearance of the Lee-Yang zeros is quite similar to the 
situation in QCD where the sign problem is present. 

The property of a first order phase transition that isolated Lee-Yang 
zeros appear periodically at $\beta_{\rm Im} \sim C(2n+1)$, where $C$ 
is the distance to the nearest Lee-Yang zero and $n$ is an integer, 
is free from the problems that arise in the infinite volume limit. 
Therefore, to investigate the pattern of the appearance of Lee-Yang 
zeros in the $(\beta_{\rm Re}, \beta_{\rm Im})$ plane is important.
For this calculation, the simulations by using high statistics data 
and large lattice size are indispensable. 
Further studies are clearly important to find the endpoint of 
the first order phase transition line in the $(T, \mu_q)$ plane.

Recently, a close relation between the strength of the sign problem 
and the position of the phase transition line for pion condensation 
in phase quenched QCD has been discussed in Ref.~\cite{spli}.  There
it has been 
found that the endpoints of the first order transition line determined in 
Ref.~\cite{FK2,FK3} are located near the phase transition line of pion 
condensation. These results may relate to our discussion given here. 

Moreover, the pathologies in the Glasgow method have been discussed 
in Ref.~\cite{Bar}. Similar problems arise also in the Glasgow method. 
It would be interesting to consider the relation between the 
pathologies in the Glasgow method and those in the Lee-Yang zero analysis.

\section*{Acknowledgments}
I would like to thank T.~Hatsuda and K.~Kanaya for 
fruitful discussions and useful comments on this manuscript. 
I also thank C.R.~Allton, M.~D\"{o}ring, S.J.~Hands, O.~Kaczmarek,
F.~Karsch, E.~Laermann, and K.~Redlich for discussions, comments 
and allowing me to use the data of the Bielefeld-Swansea collaboration
partly published in Ref.~\cite{us05}. 
The data given by QCDPAX in Ref.~\cite{QCDPAX} are used for 
the analysis in Sec.~\ref{sec:pureQCD}. 
This work is supported by BMBF under grant No.06BI102.

\begin{table}[thb]
\caption{Positions of Lee-Yang zeros for the SU(3) pure gauge theory}
\label{tab:flyz}
\begin{center}
\begin{tabular}{cccc}
Lattice size & & $\beta_{\rm Re}$ & $\beta_{\rm Im}$ \\
\hline
$12^2 \times 24 \times 4$ & 1st zero & 5.69178(23) & 0.01216(17) \\
$24^2 \times 36 \times 4$ & 1st zero & 5.69252(5) & 0.00212(2) \\
$24^2 \times 36 \times 4$ & 2nd zero & 5.69309(7) & 0.00556(7) \\
$36^2 \times 48 \times 6$ & 1st zero & 5.89411(10) & 0.00434(8) \\
\end{tabular}
\end{center}
\end{table}

\begin{table}[thb]
\caption{Values of $\mu_q/T \equiv N_{\tau} \mu$ at which 
$\langle \cos \theta \rangle =0.1, 0.01$ and $0.$ 
$N_{\rm site} =16^3 \times 4.$}
\label{tab:mu}
\begin{center}
\begin{tabular}{cccc}
$T/T_0$ & 
$\langle \cos \theta \rangle =0.1$ & 
$\langle \cos \theta \rangle =0.01$ & 
$\langle \cos \theta \rangle =0.0$ \\
\hline
0.90 & 0.70(2) & 1.0(2) & 1.2(2) \\
0.96 & 0.80(2) & 1.1(1) & 1.1(1) \\
1.00 & 0.87(2) & 1.9(2) & 2.3(4) \\
1.02 & 0.96(3) & 2.2(12) & 2.3(1) \\
1.07 & 1.13(3) & 1.8(4) & 2.0(2) \\
\end{tabular}
\end{center}
\end{table}

\end{document}